\documentclass[10pt,a4paper]{article} 
\title{\bf Entropic Distance for Nonlinear Master Equation}

\author{
Tam\'as S. Bir\'o\footnote{A talk based on this work was presented by T.S.Bir\'o at the BGL 2017 Gy\"ongy\"os, Hungary},  Andr\'as Telcs 
\\ H.A.S. Wigner Research Centre for Physics, Budapest, Hungary 
\and
Zolt\'an N\'eda
\\  University  Babes-Bolyai, Cluj, Romania; zneda@phys.ubbcluj.ro
}

\usepackage{graphics}
\usepackage{graphicx}
\usepackage{xcolor}

\usepackage[thicklines]{cancel}

\usepackage{amssymb}

\usepackage{bm}

\begin{document}

\maketitle

% Abstract (Do not use inserted blank lines, i.e. \\) 
\abstract{
More and more works deal with statistical systems far from equilibrium, dominated by unidirectional stochastic processes
augmented by rare resets. We analyze the construction of the entropic distance measure appropriate for such a dynamics.
We demonstrate that a power-like nonlinearity in the state probability in the master equation naturally leads to
the Tsallis (Havrda--Charv\'at, Acz\'el--Dar\'oczy) $q$-entropy formula in the context of seeking for the maximal entropy 
state at stationarity.
A few possible applications of a certain simple and linear master equation to phenomena studied in statistical physics
are listed at the end.
}

% Keywords
\vspace{1mm}
Keywords: {q-entropy; entropic distance; Matthew principle}
\vspace{1mm}

%%%% ~/tex/NETWORK/tsbiro_BGL17.tex

%  2017 JUL 17		% for a talk in Gyongyos

%%%%%%%%%%%%% DEFINE CUSTOM COLORNAMES %%%%%%%%%%
\definecolor{Pergamen}{RGB}{235,225,200}
\definecolor{LightGray}{RGB}{235,235,230}
\definecolor{PaleBlue}{RGB}{190,210,255}
\definecolor{DarkGreen}{RGB}{0,80,20}
\definecolor{SoftRed}{RGB}{255,220,170}
\definecolor{DarkBlue}{RGB}{0,10,80}
\definecolor{Mahogany}{RGB}{0.75, 0.25, 0.0}
\definecolor{goldenrod}{rgb}{0.72, 0.53, 0.04}
\definecolor{BrickRed}{rgb}{0.8, 0.25, 0.33}
\definecolor{RawSienna}{cmyk}{0,0.72,1,0.45}
\definecolor{RubineRed}{rgb}{0.88, 0.07, 0.37}

\newcommand{\vs}{\vspace{3mm}}

\newcommand{\be}{\begin{equation}}
\newcommand{\ee}[1]{\label{#1} \end{equation}}
\newcommand{\ba}{\begin{eqnarray}}
\newcommand{\ea}[1]{\label{#1} \end{eqnarray}}
\newcommand{\nl}{\nonumber \\}
\newcommand{\re}[1]{(\ref{#1})}
\newcommand{\spr}[2]{\vec{#1}\cdot\vec{#2}}
\newcommand{\ave}{\overline{u}}
\newcommand{\ve}[1]{\left\vert #1  \right\vert}
\newcommand{\exv}[1]{ \left\langle {#1} \right\rangle}

\newcommand{\pd}[2]{ \frac{\partial #1}{\partial #2}}
\newcommand{\pt}[2]{ \frac{{\rm d} #1}{{\rm d} #2}}
\newcommand{\pv}[2]{ \frac{\delta #1}{\delta #2}}

\newcommand{\grad}{{\vec{\nabla}}}

\newcommand{\ead}[1]{ {\rm e}^{#1}}
\newcommand{\infi}{ \int_{0}^{\infty}\limits\!}
\newcommand{\prodj}{ \prod_{j=1}^{n}\limits }
\newcommand{\sumn}{ \sum_{n=0}^{\infty}\limits\! }
\newcommand{\sumnat}{ \sum_{n=1}^{\infty}\limits\! }

\section{Definition and Properties of Entropic Distance}

Entropic distance, more properly called ''entropic divergence'', is traditionally interpreted as a relative entropy,
as a difference between entropies with a prior condition and without \cite{MANKE}. It is also the Boltzmann--Shannon entropy
of a distribution relative to another \cite{KULLBACK}. Looking at this construction, however, from the viewpoint of
a generalized entropy \cite{qCSISZAR}, the simple difference or logarithm of a ratio cannot be hold as a definition any more.

Instead, in this paper, we explore a reverse engineering concept: seeking for an entropic divergence formula at the
first place, which is subject to some wanted properties, we consider entropy as a derived quantity.
More precisely we seek for entropic divergence formulas appropriate to a given stochastic dynamics, 
shrinking during the approach to a stationary distribution, whenever it exists, and establish the entropy
formula from this distance to the uniform distribution. By doing so we serve two goals: i) having constructed
a non-negative entropic distance we derive an entropy formula which is maximal for the uniform distribution,
and ii) we come as near as possible to the classical difference formula for the relative entropy.

%%%%%%%%%%%%%%%%%%%%%% entropic distance evolution %%%%%%%%%%%%%%

We start our discussion by contrasting the definition of the metric distance, knwon from geometry, to
the basic properties of an entropic distance.
The metric distance possesses the following properties:
\begin{enumerate}
 \item $\rho(P,Q) \ge 0$ for a pair of points $P$ and $Q$,
 \item $\rho(P,Q)=0$ only for $P=Q$,
 \item $\rho(P,Q)=\rho(Q,P)$ symmetric measure,
 \item $\rho(P,Q) \le  \rho(P,R) + \rho(R,Q)$, the triangle inequality in elliptic spaces.
\end{enumerate}
The entropic divergence on the other hand is neither necessarily symmetric, nor can satisfy
a triangle inequality. On the other hand it is subject to the second law of thermodynamics,
distinguishing the time arrow from the past to the future. We require for a real functional,
$\rho[P,Q]$, depending on the distributions $P_n$ and $Q_n$, the followings to hold:
\begin{enumerate}
 \item $\rho[P,Q] \ge 0$ for a pair of distributions $P_n$ and $Q_n$,
\vs \item $\rho[P,Q]=0$ only if the distributions coincide $P_n=Q_n$,
\vs \item $\pt{}{t} \rho[P,Q] \le 0$ if $Q_n$ is the stationary distribution,
\vs  \item $\pt{}{t} \rho[P,Q] \: = \: 0$ only for $P_n=Q_n$, i.e. the stationary distribution is unique.
\end{enumerate}
Although this definition is not symmetric in the handling of the normalized distributions
$P_n$ and $Q_n$, it is an easy task to consider the symmetrized version,
${s}[P,Q] \: \equiv \: \rho[P,Q] \, + \, \rho[Q,P]$.
This symmetrized entropic divergence inherits some properties from the fiducial construction.
Considering a scaling trace form entropic divergence, $\rho[P,Q] = \sum_n\limits \sigma(\xi_n) \, Q_n$ 
with $\xi_n=P_n/Q_n$, to begin with, we identify the following symmetrized kernel function:
\be
 \mathfrak{s}(\xi) \: := \: \sigma(\xi) + \xi \, \sigma(1/\xi).
\ee{SYMDIV}
The only constraint is to start with a core function, $\sigma(\xi)$ with a definite concavity.
Jensen inequality tells for $\sigma^{\prime\prime} > 0$ that
\be
 \sum_n\limits \sigma(\xi_n) \, Q_n \: \ge \: 
 \sigma\left(\sum_n\limits \xi_n Q_n \right) 
 = \sigma\left( \sum_n\limits  P_n \right)= \sigma(1).
\ee{JENSENSIGMA}
For satisfying property 1 and 2 one simply sets {${ \sigma(1)=0}$}. Interesting enough, but this
setting suffices also for the satisfaction of the second law of thermodynamics, formulated above as
further constraints 3 and 4.
As a consequence of the symmetrization {it also follows that ${ \mathfrak{s}(1)=0}$ and $\mathfrak{s}^{\prime\prime} > 0$.}

The {symmetrized entropic divergence} shows some {new, emergent properties}.
We list its derivatives as follows:
\ba
 \mathfrak{s}(\xi) &=& \sigma(\xi) + \xi \, \sigma(1/\xi)
\nl
 \mathfrak{s}^{\prime}(\xi) &=& \sigma^{\prime}(\xi) + \sigma(1/\xi) - \frac{1}{\xi} \sigma^{\prime}(1/\xi)
\nl
 \mathfrak{s}^{\prime\prime}(\xi) &=& \sigma^{\prime\prime}(\xi) 
  - \cancel{\frac{1}{\xi^2} \sigma^{\prime}(1/\xi)} + \cancel{\frac{1}{\xi^2} \sigma^{\prime}(1/\xi)}
 + \frac{1}{\xi^3} \sigma^{\prime\prime}(1/\xi).
\ea{SDERIVS}
The consequences, listed below, can be derived from these general relations:
\begin{enumerate}
  \item \quad $\mathfrak{s}(1) = 2 \, \sigma(1) = 0$,
  \item \quad $\mathfrak{s}^{\prime}(1) = \sigma(1) = 0$,
  \item \quad $\mathfrak{s}^{\prime\prime} > 0  \quad \Rightarrow  \quad \xi_m=1$ is a minimum,
  \item \quad {$\mathfrak{s}(\xi) \ge 0$}.
\end{enumerate}
In this way the kernel function and hence each summand in the symmetrized entropic divergence formula is
non-negative, not only the total sum.

%%%%%%%%%%%%%%%%%%%% DISCRETE SUM CASE FOR GENERAL MASTER EQ %%%%%%%%%%%%

\section{Entropic distance evolution due to linear stochastic dynamics}

Now we study properties 3 and 4, by evaluating the rate of change of the entropic divergence in time.
This change is based on the dynamics (time evolution) of the evolving distribution, $P_n(t)$,
while the targeted stationary distribution, $Q_n$ is by definition time independent.
First we consider a class of stochastic evolutions governed by differential equations for
$\dot{P}_n(t) \equiv \pt{P_n}{t}$, linear in the distribution, $P_n(t)$ \cite{BIROEPJA}.
We consider the trace form $\rho[P,Q]=\sum_n\limits Q_n \, \sigma\!\left(\frac{P_n}{Q_n}\right)$ 
and the background master equation
\be
\dot{P}_n \: = \: \sum_m\limits \left(w_{nm}P_m - w_{mn}P_n \right).
\ee{LINPMASTER}
The antisymmetrized sum in the above equation is merely to ensure the conservation of the norm,
$\sum_n\limits P_n = 1$, during the time evolution.
Using again the notation $\xi_n=P_n/Q_n$ we obtain
\be
 \dot{\rho} \: = \: \sum_n \sigma^{\prime}(\xi_n) \, \dot{P}_n \: = \:
 \sum_{n,m}\limits \sigma^{\prime}(\xi_n) \, \left(w_{nm} \, \xi_mQ_m -w_{mn}\, \xi_n Q_n  \right).
\ee{RHODOTMASTER}
The basic trick is to apply the splitting $\xi_m=\xi_n + (\xi_m-\xi_n)$ to get
\be
 \dot{\rho} \: = \: \sum_n\limits  \sigma^{\prime}(\xi_n) \, \xi_n 
 \, \cancel{ \sum_m\limits \left( w_{nm} \, Q_m - w_{mn} \, Q_n \right) }
%\nl
  +  \sum_{n,m}\limits \sigma^{\prime}(\xi_n)(\xi_m-\xi_n) w_{nm} \, Q_m.
\ee{RHODOTCANCELFIRST}
Here the sum in the first term vanishes due to the very definition of the stationary distribution, $Q_n$.
For estimating the remaining term we utilize the Taylor series remainder theorem in the Lagrange form.
We recall the Taylor expansion of the kernel function $\sigma(\xi)$,
\be
  \sigma(\xi_m) \: = \: \sigma(\xi_n) + {\sigma^{\prime}(\xi_n) (\xi_m-\xi_n)}
  + \frac{1}{2} \sigma^{\prime\prime}(c_{mn}) (\xi_n-\xi_m)^2,
\ee{TAYLOR}
with $c_{mn} \in [\xi_m,\xi_n] $. Here the first derivative term has occurred in eq.(\ref{RHODOTCANCELFIRST}).
This construction delivers
\be
 \dot{\rho} \: = \: \cancel{ \sum_{n,m}\limits \left[\sigma(\xi_m) - \sigma(\xi_n) \right] w_{nm} Q_m }
 - \frac{1}{2} \sum_{n,m}\limits \sigma^{\prime\prime}(c_{mn}) \left(\xi_m-\xi_n \right)^2 w_{nm}Q_m.
\ee{RHODOTSECOND}
Here the first sum vanishes again: after exchanging the indices $m$ and $n$ in the first summand,
the result is proportional to the total balance expression, which is zero for the stationary distribution.
With positive transition rates, $w_{nm} > 0$ the approach to stationary distribution,
{${\dot{\rho} \le 0}$ {is hence proven for all} ${\sigma^{\prime\prime}>0}$.}
We note that we never used the detailed balance condition for the transition rates, only the vanishing of the
total balance, which defines the stationary distribution.

This proof, without recalling the detailed balance condition as Boltzmann's famous H-theorem did, 
is quite general. Any core function with positive second derivative and the scaling trace form
co-act to ensure the correct change in time. By using the traditional choice,
\, $\sigma(\xi)=-\ln\xi$, \, we have \, $\sigma^{\prime} = - 1/\xi$ \, and 
\, $\sigma^{\prime\prime}(\xi)= 1/\xi^2 > 0$, satisfying indeed all requirements.
The integrated entropic divergence formula (no symmetrization) in this case  
is given as the {{ Kullback--Leibler divergence }}:
\be
     \rho[P,Q] \: = \: \sum_n\limits Q_n \, \ln \, \frac{Q_n}{P_n}.
\ee{RHOLOG}
There is a rationale behind using the logarithm function. It is the only one being additive
for the product form of its argument, mapping factorizing and hence statistically independent
distributions to an additive entropic divergence kernel:
{ For $P_n^{(12)}=P_n^{(1)}P_n^{(2)}$ also $Q_n^{(12)}=Q_n^{(1)}Q_n^{(2)}$ 
therefore we have $\xi_n^{(12)}=\xi_n^{(1)}\xi_n^{(2)}$.  
\quad Aiming at $\sigma(\xi^{(12)})=\sigma(\xi^{(1)})+\sigma(\xi^{(2)})$, 
the solution is $\sigma(\xi)=\alpha\ln\xi$.
For $\sigma^{\prime\prime}>0$ it must be $\alpha<0$, so  without restricting generality
one chooses $\alpha=-1$.  } 

Finally we would like to treat this entropic divergence as an entropy difference.
This is achieved when comparing the stationary distribution to the uniform 
distribution, $U_n=1/W, n=1,2,\ldots W$.
Using the above Kullback--Leibler divergence formula one easily derives
\be
 \rho[U,Q] \: = \: \sum_{n=1}^W\limits Q_n \ln(WQ_n) \: = \: 
 \ln W + \sum_n\limits Q_n \ln Q_n
 \: = \: S_{BG}[U] - S_{BG}[Q]
\ee{DISTASDIFF2}
with 
\be
 S_{BG}[Q] \: = \: - \sum_n\limits Q_n \, \ln Q_n,
\ee{BGENTROPYFORMULA}
being the Boltzmann--Gibbs--Planck--Shannon entropy formula.
{From the Jensen inequality it follows $\rho[U,Q]\ge 0$ , 
   so $S_{BG}[U] \ge S_{BG}[Q]$. }

%%%%%%%%%%%%% NONLINEAR %%%%%%%%%%%%%%%%

\section{Entropic divergence evolution for nonlinear master equations}

Detailed balance is also not needed for a more general dynamics.
We consider Markovian dynamics, with a master equation nonlinear in the 
distribution, $P_n$, as
\be
 \dot{P}_n \: = \: \sum_m\limits \left[ w_{nm} \, a(P_m) - w_{mn} \, a(P_n) \right].
\ee{GENDYN}
The stationarity condition defines
\be
 0 \: = \: \sum_m\limits \left[ w_{nm} \, a(Q_m) - w_{mn} \, a(Q_n) \right].
\ee{GENSTAT}
The entropic distance formula is sought for in the trace form (but this time without the scaling assumption):
\be
 \rho[P,Q] \: = \: \sum_n \sigma(P_n,Q_n),
\ee{GENENETDIST}
the dependence on $Q_n$ is fixed by $\rho[Q,Q]=0$.
The change of the entropic divergence in this case is given by
\be
\dot{\rho} \: = \: \sum_{m,n}\limits \pd{\sigma}{P_n} %\sigma^{\prime}_n 
  \left[ w_{nm} \, a(Q_m) \xi_m - w_{mn} \, a(Q_n) \xi_n \right]
\ee{GENRHODOT}
with $\xi_n \: := \: a(P_n)/a(Q_n)$.
We again put $\xi_m=\xi_n + (\xi_m-\xi_n)$ in the first summand:
\be
\dot{\rho} \:  =  \: \sum_{n}\limits \pd{\sigma}{P_n} \, \xi_n \, 
\cancel{\sum_m\limits \left[ w_{nm} \, a(Q_m)  - w_{mn} \, a(Q_n) \right]}
\:  +  \: \sum_{n,m}\limits \pd{\sigma}{P_n} w_{nm} \, a(Q_m) \left( \xi_m-\xi_n \right)
\ee{GENRHODOT2}
In order to use the remainder theorem one has to identify
\be
 \pd{\sigma}{P_n}  \: = \: \kappa^{\prime} (\xi_n) \: =  \: \kappa^{\prime}\left(\frac{a(P_n)}{a(Q_n)} \right).
\ee{SPECIALSIGMAPRIME}
This ensures $\dot{\rho} < 0$ for any $\kappa^{\prime\prime}>0$ and $P \ne Q$.

We examine the example of the $q$--Kullback--Leibler or R\'enyi divergence.
Starting with the classical logarithmic kernel, $\kappa(\xi)=-\ln\xi$, 
we have $\kappa^{\prime\prime}(\xi)=1/\xi^2 > 0$.
Now having a nonlinear stochastic dynamics, $a(P)=P^{q}$,
the integrated entropic divergence formula (without symmetrization) delivers 
the Tsallis divergence \cite{qTSALLIS,qHAVRDA,qDAROCZY}, 
\be
 \pd{\sigma}{P_n} \: = \: - \frac{Q_n^{q}}{P_n^{q}}, \qquad \Rightarrow \qquad
 \rho[P,Q] \: = \: \sum_n\limits Q_n  \,  {\ln}_{q} \frac{Q_n}{P_n} .
\ee{SPECIALSIGMATSALLIS}
with
\be
 \ln_{q}(x) = \frac{1-x^{q-1}}{1-q}
\ee{LAMBDALOG}
being the so called deformed logarithm with the real parameter $q$.

We again would like to interpret this entropic divergence as entropy difference.
The entropic divergence of the stationary distribution from the uniform distribution $U_n=1/W, n=1,2,\ldots W$
is given by:
\be
 \rho[U,Q] \: = \: \sum_{n=1}^W\limits \frac{Q_n}{1-q}\left[ 1 - (WQ_n)^{q-1}\right] 
 \: = \: W^{q-1} \, \left( S_{T}[U] - S_{T}[Q] \right).
\ee{DISTASDIFF}
with $S_{T}$ being the Tsallis entropy formula:
\be
 S_{T} [Q] \: = \: \frac{1}{1-q} \sum_n\limits (Q_n^q-Q_n) \: = \: - \sum_n\limits Q_n \, {\ln_q} (Q_n).
\ee{SQTSALLIS}
From the Jensen inequality it follows $\rho[U,Q]\ge 0$ , 
so $S_{T}[U] \ge S_{T}[Q]$, i.e. the Tsallis entropy formula is also maximal for the
uniform distribution.  The factor $W^{q-1}$ signifies non-extensivity, a dependence on the number of states
in the relation between the entropic divergence and the relative Tsallis entropy.

%%%%%%%%%%%%%%%%%%%%%%%%%%%% DYNAMICAL MODELS %%%%%%%%%%%%%%%%%%

\section{Master equation for unidirectional growth and reset}

With the particular choice of the transition rates, $w_{nm}=\mu_m \delta_{n-1,m} + \gamma_m \delta_{n,0} $,
one describes a local growth process augmented with direct resetting transitions from any state to the
ground state labelled by the index zero \cite{BIRONEDA}. The corresponding master equation
\be
	\dot{P_n} \: = \: \mu_{n-1} P_{n-1} \, - \, \left(\mu_n + \gamma_n \right) \, P_n
\ee{EXPSCALEDMASTER}
is terminated at $n=1$ and the equation for the $n=0$ state takes care of the normalization conservation:
\be
\dot{P}_0 \: = \: \sum_{n=1}^{\infty}\limits \gamma_n P_n \,   - \, \mu_0 P_0. 
\ee{ZEROPROBEQ}
For the stationary distribution one obtains
\be
 Q_n \: = \: \frac{\mu_{n-1}}{\mu_n+\gamma_n} \, Q_{n-1} \: = \: \cdots \: = \:
 \frac{\mu_0Q_0}{\mu_n} \, \prod_{j=1}^{n}\limits \left(1+\frac{\gamma_j}{\mu_j} \right)^{-1},
\ee{EXPSCALEDMASTERSTAC}
and $Q_0$ has to be obtained from the normalization.
Table \ref{TABLEONE} summarizes some well known probability density functions, PDF-s,
which emerge as stationary distribution to this
simplified stochastic dynamics upon different choices of the growth and reset rates
$\mu_n$ and $\gamma_n$. In the continuous limit we obtain
\be
\pd{}{t} P(x,t) \: = \:  - \pd{}{x} \left(\mu(x) \, P(x,t) \right) - \gamma(x) P(x,t).
\ee{CONTIMASTER}
with the stationary distribution
\be
Q(x) \: = \:  \frac{K}{\mu(x)} \, \ead{-\int_0^x\limits \frac{\gamma(u)}{\mu(u)} \, du}.
\ee{CONTISTATIONARY}
\begin{table}[h]
\caption{\label{TABLEONE} Summary of rates and stationary PDF-s.}
\begin{center}
\begin{tabular}{||c|c|c||}
\hline \hline
$\gamma_n$, $\gamma(x)$ & $\mu_n$, $\mu(x)$  & $Q_n$, $Q(x)$  \\ \hline \hline
 const & const  & geometrical $\to$  exponential \\ \hline
 const & linear  & Waring $\to$ Tsallis/Pareto\\ \hline
 const & sublinear power  & Weibull\\ \hline
 const & quadratic polynomial & Pearson \\ \hline
 const & exp  & Gompertz \\ \hline 
 $\ln (x/a)$ & $\alpha x$ & Log-Normal \\ \hline
  linear  & const & Gauss \\ \hline
 $\alpha(ax-c)$ & $\alpha x$ & Gamma \\ \hline
\hline
\end{tabular}
\end{center}
\end{table}

Finally we derive a bound for the entropy production in the continuous model of unidirectional growth
with resetting.

%%%%%%%% Continuous %%%%%%%%
First we study the time evolution of the ratio, $\xi(t,x)=P(x,t)/Q(x)$.
Using $P=\xi Q$ we get from eq.(\ref{CONTIMASTER}):
\be
Q \, \pd{\xi}{t} \: = \: - \xi \, 
\cancel{{\partial (\mu Q)} \over {\partial x} } 
- \mu Q \, \pd{\xi}{x} - \cancel{ \gamma \, Q } \, \xi.
\ee{CONTIMASTERWITHRATIO1}
Using the same eq. for stationary $Q(x)$ and dividing by $Q$ we obtain
\be
 \pd{\xi}{t} \: = \: - \mu(x) \, \pd{\xi}{x}.
\ee{CONTIMASTRATIO1}
Now we turn to the evolution of the entropic divergence, 
\be
 \rho(t) \: \equiv \:  \infi \mathfrak{s}(\xi(t,x)) \, Q(x) dx,  
\ee{RHODEF}
With the symmetrized kernel, $\mathfrak{s}(\xi) = \sigma_{\mathrm{div}}(\xi) + \xi \, \sigma_{\mathrm{div}}(1/\xi) \ge 0$,
one gets using \quad $\pd{\mathfrak{s}}{t}=-{\mu(x)}\pd{\mathfrak{s}}{x}$  the following
distance evolution, considering the boundary condition $\xi(t,0)=1$  and  $\mathfrak{s}(1) = 0$:
\be
 \pt{\rho}{t} \: = \: 
 - \infi \mathfrak{s}(\xi(t,x)) \, Q(x) \, \gamma(x) dx  
\ee{CONSOLVESR}
We note that for the Kullback--Leibler divergence the following
symmetrized kernel function has to be used: {$\sigma(\xi)=-\ln \xi$ } leads to
{$\mathfrak{s}(\xi)=(\xi-1) \, \ln \xi $}  
and in this way ensures $\pt{\rho}{t} \le 0$. 

%%%%%%%%%%%% JENSEN BOUND FOR ENTROPY RATE %%%%%%%%%%%%%%%
In order to obtain a lower bound for the speed of the approach to stationarity, we use again the
{Jensen inequality} {for $\mathfrak{s}(\xi)$}:
\be
 \int \! p(x) \, \mathfrak{s}(\xi(x))\, dx \: \ge \: \mathfrak{s}\left( \int \! p(x) \, \xi(x) \, dx \right)
\ee{CONT_JENSEN}
with any arbitrary $p(x) \ge 0$ satisfying $\int\! p(x) \, dx = 1$.
For pour purpose we choose $p(x)=\gamma(x)Q(x)/\int\! \gamma Q \, dx$.
This leads to the following result:
\be
 \pt{\rho}{t} \: \le \: 
 - \exv{\gamma}_{\infty} \cdot \mathfrak{s}\left(\frac{\exv{\gamma}_{\: t}}{\exv{\gamma}_{\infty}} \right)  
 \: = \: \Big[ \exv{\gamma}_{\infty} - \exv{\gamma}_{\: t} \Big] \, \cdot \, \ln \frac{\exv{\gamma}_{\: t}}{\exv{\gamma}_{\infty}}.
%\bm{\mu(0)Q(0)\mathfrak{s}(\xi(t,0)) - \infi \mathfrak{s}(\xi(t,x)) \, Q(x) \, \gamma(x) dx  }%
\ee{CONSOLVESR2}
Note that the controlling quantity is actually the expectation value
of the resetting rate,  $\int\! p(x) \xi(x) \, dx = \int\! \gamma P \, dx = \exv{\gamma}_t$.
Since $\mathfrak{s}(\xi)$ reaches its minimum with the value zero only at the argument $1$,
the entropic divergence $\rho(t)$ stops changing only if the stationary distribution is achieved.
In all other cases it shrinks.

\section{Summary}
%%%%%%%%%%%%%%%%%%% APPLICATIONS %%%%%%%%%%%%%%%%%

Summarizing, in this paper we have presented a construction strategy for the entropic distance
formula, designed to shrink for a given wide class of stochastic dynamics. The very entropy
formula was then derived from inspecting this distance between the uniform distribution and
the stationary PDF of the corresponding master equation. In this way for linear master
equations the well-known Kullback--Leibler definition arises, while for nonlinear dependence
on the occupation probabilities one always arrives at an accordingly modified expression.
In particular for a general power-like dependence the Tsallis $q$-entropy occurs as the
''natural'' relative entropy interpretation of the proper entropic divergence.
In the continuous version of the growth and reset master equation, a dissipative probability
flow supported with an inflow at the boundary, a lower bound was given for the shrinking
speed of the symmetrized entropic divergence using the Jensen inequality.

To finish this paper we would like to make some remarks on real world applications of the above
discussed  mathematical treatment.
Among possible applications of the growth and resetting model 
we mention the network degree distributions showing exponential behavior
for constant rates and a Tsallis--Pareto distribution \cite{PARETO} 
(in the discrete version a Waring distribution \cite{WARING,NETWARING})
for having a linear preference in the growth rate, $\mu_n=\alpha(n+b)$. For high energy particle abundance
(hadron multiplicity) distributions the negative binomial PDF is an excellent approximation \cite{GBIRO}, 
when both rates $\mu$ and $\gamma$ are linear functions of the state label.
For middle and small settlement size distributions
a log-normal PDF arise, achievable with linear growth rate, $\mu(x)$ and a logarithmic reset rate, $\gamma(x) \sim \ln x$.
Citations of scientific papers and Facebook shares and likes also follow a scaling Tsallis--Pareto distribution
\cite{SCIENTO,VARGANEDABIRO},
characteristic to constant resetting and linear growth rates. While wealth seems to be distributed according to
a Pareto-law tail, the middle class incomes rather show a gamma distribution, stemming from linear reset and
growth rates. For a review of such applications see our forthcoming work.

%%%%%%%%%%%%%% ACKNOWLEDGEMENT %%%%%%%%%%
%\acknowledgments{
{\bf Acknowledgment:}
This work has been supported by the Hungarian National Bureau for Research Development and Innovation,
NKFIH under project Nr. K 123815. 
%}

%%%%%%%%%%%%%%%%%% REFERENCES %%%%%%%%%%%%%%%%

%%%%%%%%%%%%%%%%%%%%%%%%%%%%%%%%%%%%%%%%%%
%\authorcontributions{
{\em Author Contributions:}

This paper is based on the talk given by T.~S.~Bir\'o at the Bolyai-Gauss-Lobachevskii conference, BGL 2017, in Gy\"ongy\"os, Hungary.
T.~S.~Bir\'o invented, suggested and worked out the unidrectional reset model and delivered the derivation of the evolution of
the entropic distance based on the stochastic master equations. A.~Telcs contributed the lower-bound estimate on the entropic
distance rate of change based on the Jensen inequality. Z.~N\'eda worked on the resetting model, its generalization for
expanding sample space processes and on collecting convincingly interpreting data for the applications.
%}

\vs
%%%%%%%%%%%%%%%%%%%%%%%%%%%%%%%%%%%%%%%%%%
%\conflictsofinterest{
The authors declare no conflict of interest.
%}

\vs
%%%%%%%%%%%%%%%%%%%%%%%%%%%%%%%%%%%%%%%%%%
%% optional
%\abbreviations{
The following abbreviations are used in this manuscript: ~\hfill~ 
 PDF -- probability density function
%}

\end{document}